\documentclass[twocolumn]{aastex63} 
\usepackage[flushleft]{threeparttable}
\usepackage{hyperref}
\usepackage{longtable}
\usepackage{tabularx}
\usepackage{amsmath,amstext}
\usepackage[figure,figure*]{hypcap}
\usepackage{newtxmath} 
\usepackage{makecell}
\usepackage{gensymb}

\shorttitle{Metal-Poor Stars Observed with SALT}
\shortauthors{Rasmussen et al.}

\submitjournal{ApJ}
\accepted{October 8, 2020}

\DeclareUnicodeCharacter{2212}{-} 
\begin{document}

\title{Metal-Poor Stars Observed with the Southern African Large Telescope} 

\correspondingauthor{Timothy C. Beers}
\email{tbeers@nd.edu}
\author{Kaitlin C. Rasmussen}
\affiliation{Department of Physics, University of Notre Dame, Notre Dame, IN 46556, USA} 
\affiliation{JINA Center for the Evolution of the Elements (JINA-CEE), USA}
\affiliation{Department of Astronomy, University of Michigan, 1085 S. University Ave., Ann Arbor, MI 48109, USA}

\author{Joseph Zepeda}
\affiliation{Department of Physics, University of Notre Dame, Notre Dame, IN 46556, USA} 
\affiliation{JINA Center for the Evolution of the Elements (JINA-CEE), USA}

\author{Timothy C. Beers} 
\affiliation{Department of Physics, University of Notre Dame, Notre Dame, IN 46556, USA} 
\affiliation{JINA Center for the Evolution of the Elements (JINA-CEE), USA}

\author{Vinicius M. Placco} 
\affiliation{Department of Physics, University of Notre Dame, Notre Dame, IN 46556, USA}  
\affiliation{JINA Center for the Evolution of the Elements (JINA-CEE), USA}
\affiliation{NSF's Optical-Infrared Astronomy Research Laboratory, Tucson, AZ 85719, USA}

\author{\'Eric Depagne} 
\affiliation{South African Astronomical Observatory (SAAO), Observatory Road Observatory Cape Town, WC 7925, South Africa}  

\author{Anna Frebel} 
\affiliation{Department of Physics and Kavli Institute for Astrophysics and Space Research, Massachusetts Institute of Technology, Cambridge, MA 02139, USA}
\affiliation{JINA Center for the Evolution of the Elements (JINA-CEE), USA}

\author{Sarah Dietz} 
\affiliation{Department of Physics, University of Notre Dame, Notre Dame, IN 46556, USA}  
\affiliation{JINA Center for the Evolution of the Elements (JINA-CEE), USA}

\author{Tilman Hartwig} 
\affiliation{Institute for Physics of Intelligence, School of Science, The University of Tokyo, Bunkyo, Tokyo 113-0033, Japan}
\affiliation{Kavli IPMU (WPI), UTIAS, The University of Tokyo, Kashiwa, Chiba 277-8583, Japan}
\affiliation{Department of Physics, School of Science, The University of Tokyo,
Bunkyo, Tokyo 113-0033, Japan}

\begin{abstract}

We present the first release of a large-scale study of relatively bright ($V < 13.5$) metal-poor stars observed with the Southern African Large Telescope (SALT), based on high-resolution spectra of 50 stars with a resolving power of $R \sim$ 40,000 and S/N $\sim$ 20 per pixel at 4300\,\AA. The elemental abundances of C, Sr, Ba, and Eu are reported, as well as several $\alpha$-elements (Mg, Ca, Sc, Ti, V) and iron-peak elements (Mn, Co, Ni, Zn). We find a diverse array of abundance patterns, including several consistent with the signatures of carbon-enhanced metal-poor CEMP-$i$ and CEMP-$r$ stars. We find that 15 of 50 (30\%) are carbon enhanced (with [C/Fe] $>$ +0.70), and that a large fraction (26 of 50,  52\%) are enhanced in $r$-process elements. Among the $r$-process-enhanced stars, five are strongly enhanced $r$-II ([Eu/Fe] $> +1.0$) stars ( two of which are newly discovered) and 21 are newly discovered moderately enhanced $r$-I ($+0.3 \leq$ [Eu/Fe] $\leq +1.0$) stars. There are eight stars in our sample that, on the basis of their abundances and kinematics, are possible members of the metal-weak thick-disk population.  We also compare  our measured abundances to progenitor-enrichment models, and find that the abundance patterns for the majority of our stars can be attributed to a single (rather than multiple) enrichment event. 

\end{abstract}


\section{Introduction}

Metal-poor stars are the oldest survivors of the earliest era of star formation after the Big Bang; they preserve material in their photospheres that has (with the exception of mass-transfer binaries) remained essentially unaltered since their birth (\citealt{Beers1992}, and references therein). This allows for the direct examination of star-formation environments that have not existed for billions of years. Their frequencies, elemental abundances, and kinematics provide insight into early star-formation channels, nucleosynthesis pathways, and Galactic assembly. As such, they are a valuable tool for studying many facets of the early Universe.

Metal-poor stars exhibit diverse elemental-abundance patterns; it is thus useful to define several sub-classes. Stars that are over-abundant in carbon relative to iron, compared to the Solar ratio ([C/Fe] \footnote{Relative abundance ratios used in this work correspond to $\textrm{[A/B]} = \log{(N_A/N_B) - \log{(N_A/N_B)}_{\odot}}$, where $N$ indicates the number density of the given species.} $> $ +0.70), are referred to as carbon-enhanced metal-poor (CEMP) stars. Stars that are also over-abundant in the neutron-capture elements are referred to as CEMP-$s$, CEMP-$i$, or CEMP-$r$, depending on their overall heavy element-abundance patterns. The CEMP-no stars exhibit under-abundances of neutron-capture elements relative to the Solar ratios. Quantitative abundance definitions can be found in Table~\ref{tab:classes}, adapted from \citet{Beers2005} and \citet{Frebel2018}. 

At very low metallicities, the presence of large amounts of carbon relative to iron is significant because it a strong indicator that a star likely formed in the early Universe through cooling channels involving CII and/or OI \citep{Bromm2003, Frebel2007b}.  This is supported by the observed increase in the cumulative fraction of CEMP stars that have not been externally enriched by a mass-transfer event from a binary companion (i.e., CEMP-no and CEMP-$r$), as demonstrated in \citet{lee2013b}, \citet{Placco2014}, and  \citet{Yoon2018}. According to the most recent work, this fraction grows steadily from $\sim$30\% at [Fe/H] $<-$2.5 to $\sim 100$\% at [Fe/H] $<-4.5$. The search for CEMP-no stars is thus important, since these stars have the greatest potential to represent bona-fide second-generation stars.


The presence of the (predominantly) $r$-process element Eu in a very metal-poor star (along with [Ba/Eu] $< 0$, to exclude possible $s$- or $i$-process contributions) indicates that an explosive progenitor event, such as a neutron star merger \citep{Lattimer1974, Arcones2007, Thielemann2017}, likely enriched the star's natal cloud with significant amounts of heavy elements beyond the iron peak. Specifically, the $r$-process-element abundance patterns (strictly speaking, the residuals from the Solar pattern after subtracting the $s$-process contribution, e.g., \citealt{Burris2000}) of such stars encode unique information on the nature of the possible  astrophysical sites and the various nucleosynthetic pathways involved in the operation of the $r$-process. 

In addition to C and Eu, several other elemental abundances can be useful tests of first-star and Galactic chemical-evolution models (e.g., \citealt{Cote2016, Cote2017}, and references therein). Measurements of $\alpha$-capture elements, such as Mg, can be used to between models for first-star nucleosynthesis mechanisms (\citealt{Yoon2016}, and references therein), as well as to separate stars with single supernova progenitors from those with multiple such progenitors \citep{Hartwig2018}. Other neutron-capture element abundances, such as Ba, can  provide insight into the $s$-process operating in asymptotic giant branch (AGB) stars and the subsequent mass transfer across a binary system \citep{Herwig2005, Bisterzo2010, Abate2015}, or in massive, rapidly rotating extremely low-metallicity stars (e.g.,\citealt{Pignatari2008}, \citealt{Meynet2010}, \citealt{Maeder2015}, \citealt{Choplin2016}, \citealt{Choplin2018}), as well as the possible operation of the $i$-process, the astrophysical site(s) for which are still under investigation \citep{Hampel2016,Denissenkov2017}.

\begin{deluxetable}{lr}
\tablecaption{\label{tab:classes} Metal-Poor Sub-class Definitions}
\tabletypesize{\scriptsize}
\tablehead{
\colhead{Sub-Classes}  & 
\colhead{Definition}}
\startdata
\multicolumn{2}{c}{$r$-process-enhanced stars}   \\
$r$-I & 0.3 $\leq$ [Eu/Fe] $\leq$ +1.0, [Ba/Eu] $<$ 0\\
$r$-II & [Eu/Fe] $>$ +1.0, [Ba/Eu] $<$ 0\\
\hline
\multicolumn{2}{c}{Carbon-enhanced metal-poor stars}   \\
CEMP & [C/Fe] $>$ +0.7\\
CEMP-$r$ &  [C/Fe] $>$ +0.7, [Eu/Fe] $>$ +1.0\\
CEMP-$s$ &  [C/Fe] $>$ +0.7, [Ba/Fe] $>$ +1.0, [Ba/Eu] $>$ +0.5\\
CEMP-$i$ & [C/Fe] $>$ +0.7, 0.0 $<$ [Ba/Eu] $<$ +0.5 
($−$1.0 $<$ [Ba/Pb] $<$ $−$0.5)\\
CEMP-no &  [C/Fe] $>$ +0.7, [Ba/Fe] $<$ 0\\
\enddata
\end{deluxetable}

\section{Target Selection and Observations} \label{sec:style}

\subsection{Target Selection}

The stars that comprise this sample are the first data release from a total of $\sim$200 metal-poor stars, selected from candidates identified during the Radial Velocity Experiment (RAVE) survey \citep{Steinmetz2006,Kunder2017}. RAVE obtained moderate-resolution (R $\sim$ 7500) spectroscopy of bright stars in the region of the Ca triplet, and derived stellar  parameters ($T_{\rm eff}$, log $g$, and [Fe/H]) and abundance estimates for a limited number of elements. Their sample of nearly 500,000 stars was selected on apparent magnitude, effectively removing the biases typically associated with searches for metal-poor stars, such as selection on metallicity itself, kinematics, coverage of a limited range of evolutionary status, or membership in a specific Galactic population. 

\begin{figure*}[ht!]
\includegraphics[trim={1cm 0cm 1cm 0cm}, scale=1]{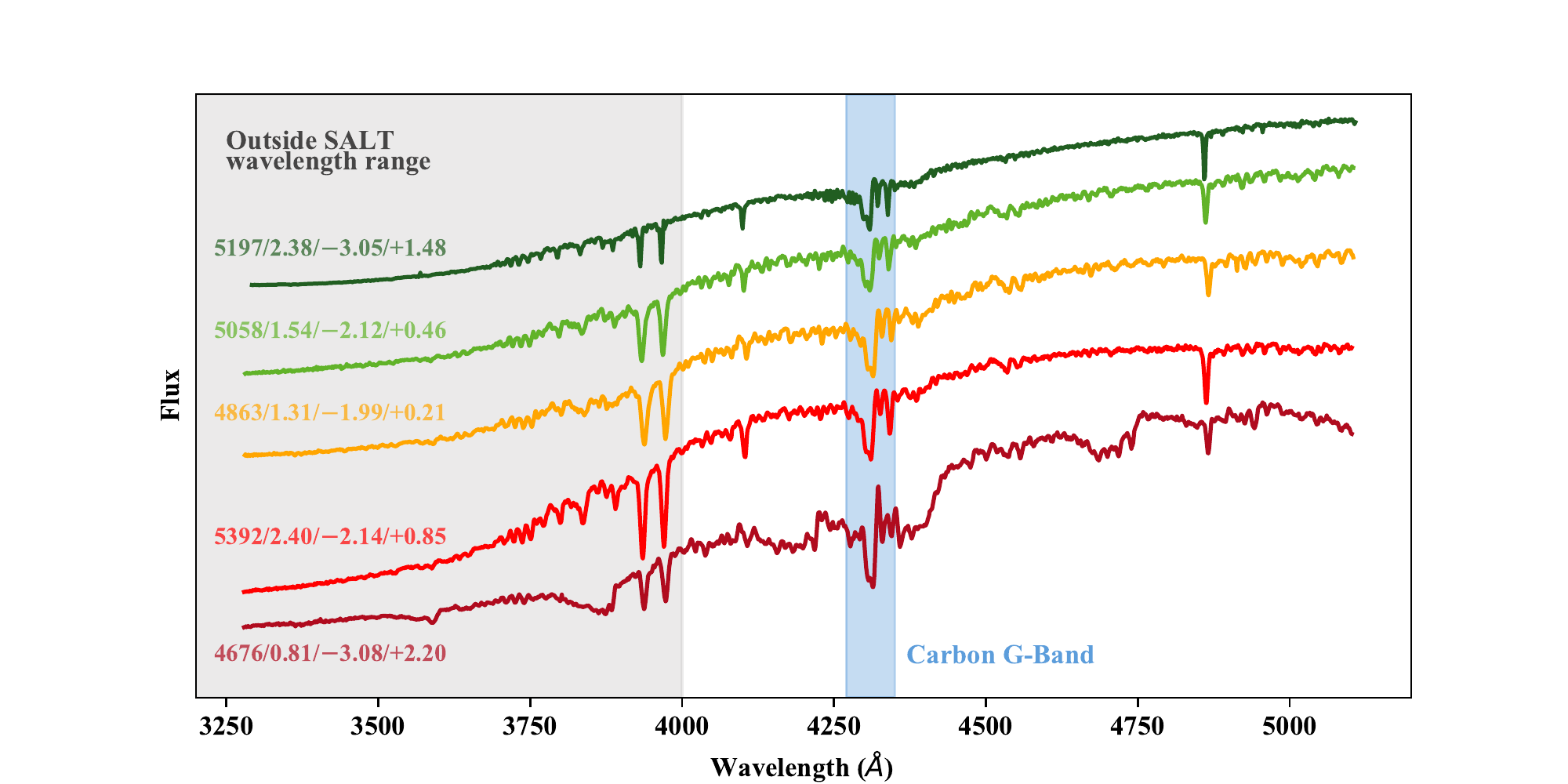}
\caption{\label{fig:mr}Medium-resolution ($R \sim$ 2000) spectra of several stars in our sample. Estimates of $T_{\rm eff}$/log $g$/[Fe/H]/[C/Fe] obtained by the n-SSPP are shown for each star:  J153830.9$-$180424 (light green), J100709.2$-$180947 (orange), J061950.0$-$531212 (red), and J044208.2$-$342114 (dark red) were taken with the EFOSC2 spectrograph at the New Technology Telescope (NTT), while J222236.0$-$013827 (dark green) was taken with the RCSPEC spectrograph at the Mayall 4m Telescope.  The gray region at the left indicates the wavelength range not covered by the high-resolution SALT spectra.  The blue-shaded region indicates the location of the CH $G$-band used by the n-SSPP to estimate [C/Fe]. The values of [C/Fe] shown have been corrected for evolutionary effects, as discussed by \citet{Placco2014}.} 
\end{figure*}

To arrive at the current sample, medium-resolution ($R \sim 2000$) spectra  of candidate  metal-poor RAVE stars were obtained (see Figure~\ref{fig:mr}), and processed by the non-SEGUE Stellar Parameter Pipeline (n-SSPP) \citep{Lee2013,Beers2017}, which compares observed spectra to synthetic spectra to predict [C/Fe], along with their stellar atmospheric parameters ($T_{\rm eff}$, $\log g$, and [Fe/H]), using a number of approaches \citep{Placco2018}. The estimates of [C/Fe] were then corrected for evolutionary effects, as discussed by \cite{Placco2014}, in order to better approximate their natal C abundances.

\begin{figure}[ht!]
\epsscale{2}
\includegraphics[trim={.75cm 0 0cm 0}, scale=0.58]{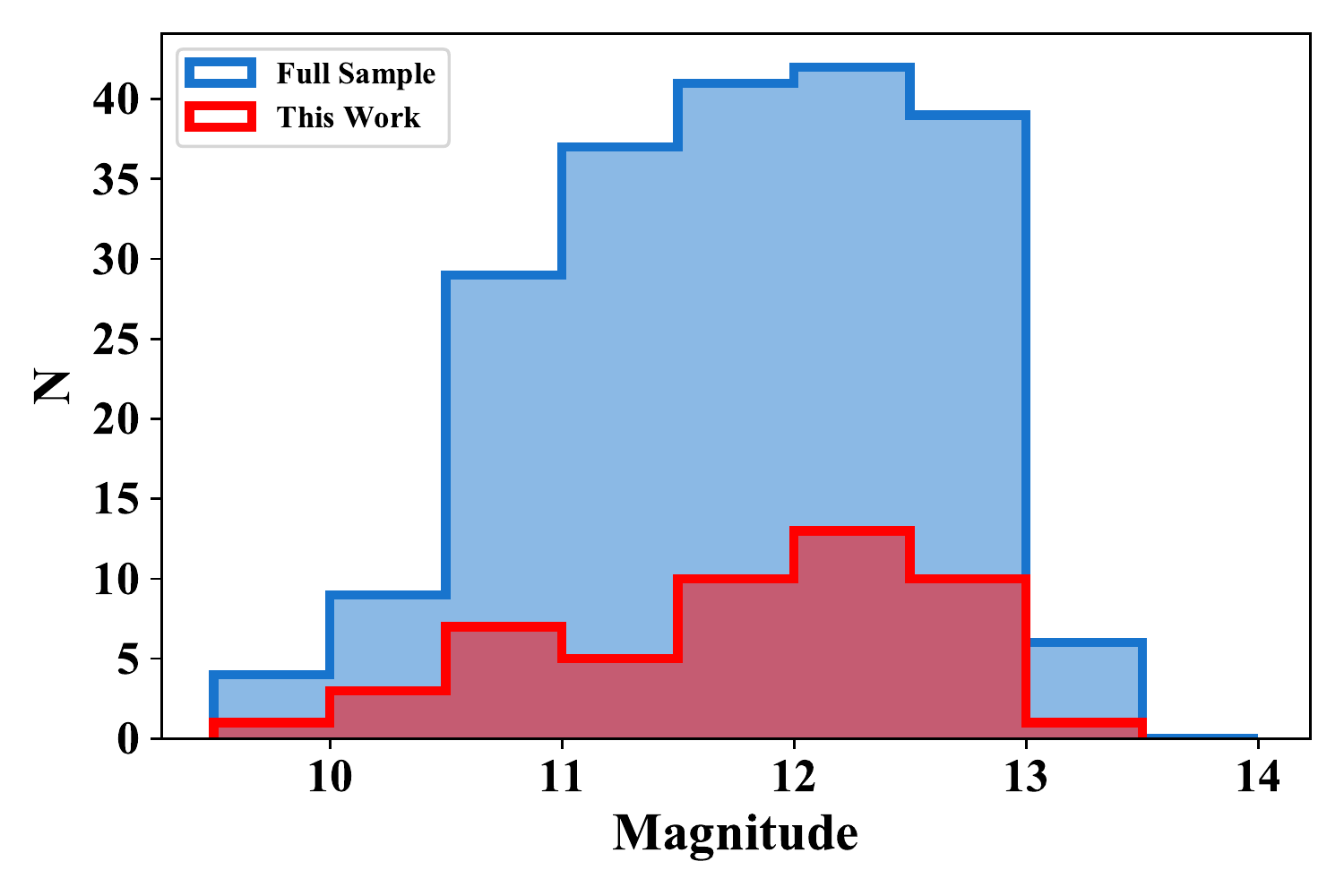}
\caption{\label{fig:vmags}Distribution of $V$ magnitudes for the present sample, as well as for the complete set of $\sim$200 stars observed with SALT on this program.} 
\end{figure}

The full sample of $\sim$200 SALT stars were selected after this step on the basis of brightness and/or their high [C/Fe] abundances, to maximize the numbers of likely CEMP stars in the sample. The sample of 50 stars reported on here were picked from the full sample as a pilot study. This pilot sample covers the same apparent magnitude range as the full $\sim$200 star sample (see Figure~\ref{fig:vmags}).

\subsection{SALT Observations}

\begin{figure*}[ht!]
\includegraphics[trim={1cm 0 1cm 0}, scale=.77]{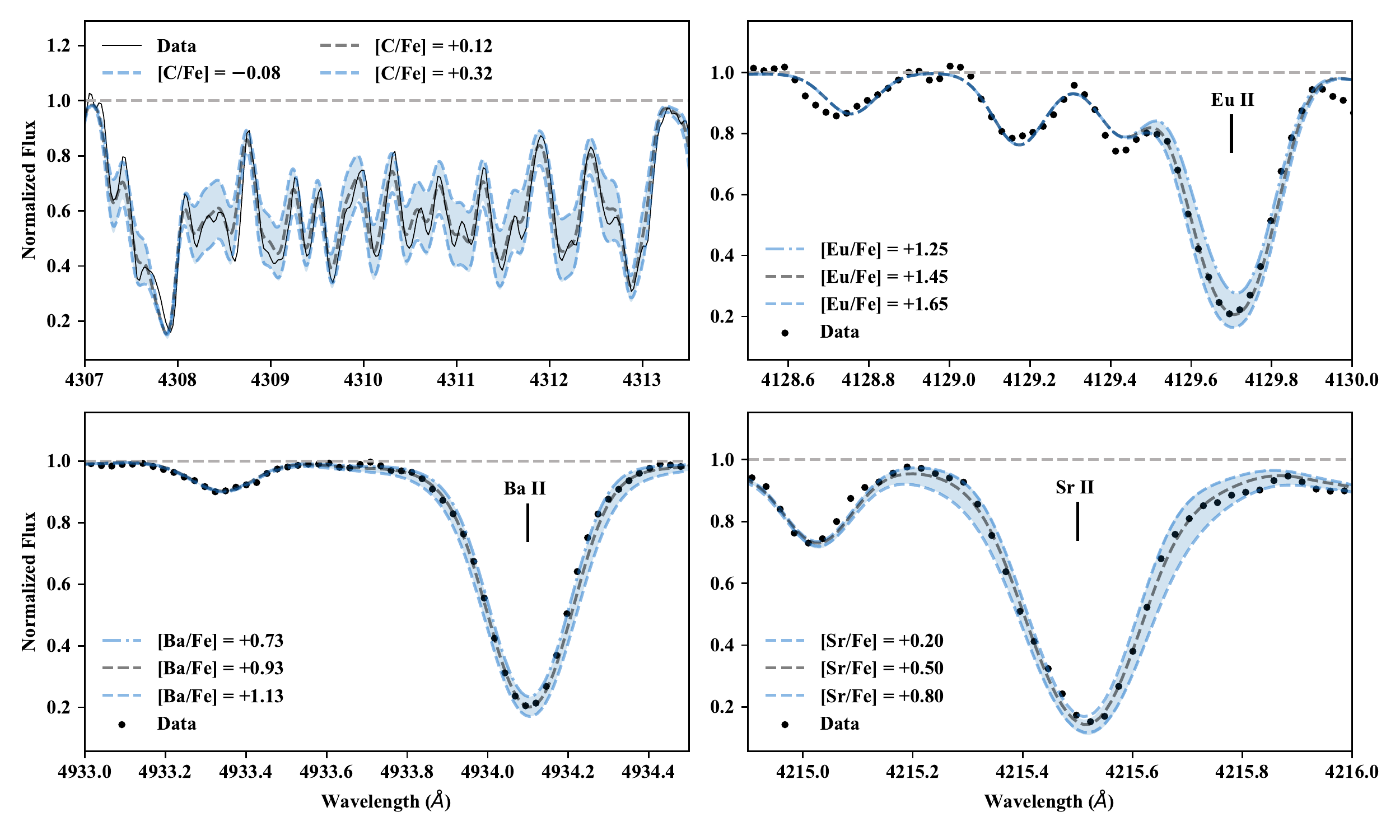}
\caption{\label{fig:hr}Comparison of high-resolution data with synthetic spectra for the elements carbon, europium, barium, and strontium, for the $r$-II star J153830.9-180424. The black dots indicate the observed spectra, while the blue-shaded region indicates the range around the best-fit values, as provided in the legend for each panel.  Top left: The CH $G$-band at $\sim  \lambda$4310\,{\AA}. Top right: The Eu\,II line at $\lambda$4129\,{\AA}. Bottom left: the Ba\,II line at $\lambda$4934\,{\AA}. Bottom right: The Sr\,II line at $\lambda$4215\,{\AA}.} 
\end{figure*}

A total of 55 hours of observing time was granted at the 11\,m Southern African Large Telescope (SALT; \citealt{Buckley2006}) for the Long Term Proposal 2017-1-MLT-012, ``Detailed Study of CEMP Stars Identified in the RAVE Survey''. The observations were carried out in service mode during four consecutive semesters, beginning in 2017 April and ending in 2019 March. A total of 223 stars were observed with the High Resolution Spectrograph (HRS; \citealt{Bramall2010,Bramall2012,Crause2014}) at a resolving power of $R \sim$ 40,000. Examples of several of the important absorption features are shown in Figure~\ref{fig:hr}.

The HRS is a dual-beam, single-object, fiber-fed e\'chelle spectrograph which utilizes a pair of fibers to simultaneously image a target and the nearby sky background. It employs Volume Phase Holographic (VPH) gratings composed of dichromated gelatin between two optical windows as cross-dispersers. HRS has three resolution settings ($R \sim$14,000, $R \sim$40,000, and $R \sim$65,000), which deliver light to one blue (370-550 nm) 2k by 4k CCD chip and one red (550-890 nm) 4k by 4k fringe-suppressing deep-depletion chip. Nominal exposure times were $\sim$2\,h, depending on the brightness of the target, leading to a typical signal-to-noise ratio (S/N) of $\sim$ 20 per pixel at 4300\,\AA. The useful wavelength range of the spectra is 3990-5560\,\AA, as the signal is dominated by the noise blueward of 3990\,\AA, and issues within our own reduction pipeline made reduction of the red wing infeasible. Further observing details can be found in Table~\ref{tab:obs}.

\bigskip
\bigskip
\bigskip
\bigskip

\section{Data Analysis and Stellar Parameters}
\subsection{Data Reduction}

Reduction of the raw spectra from HRS was carried out with custom-made code designed for SALT spectra utilizing IRAF routines \citep{Tody1986,Tody1993}. The packages \texttt{noao, imred, cddred}, and \texttt{echelle} were employed, specifically, using the tasks \textit{ccdproc, imcombine, apflatten, apscatter}, and \textit{apall}. 

Spectra were normalized, stitched together, and radial-velocity corrected with the software Spectroscopy Made Hard (SMH; \citealt{Casey2014}), which is a python wrapper of the FORTRAN-based MOOG routine \citep{Sneden1973,Sobeck2011}. Atomic data for the absorption lines used for equivalent-width determination are listed in the Appendix.

\subsection{Stellar Parameters}

\begin{figure}[h!]
\epsscale{2}
\includegraphics[trim={1cm 0cm 1cm 0cm}, scale=0.65]{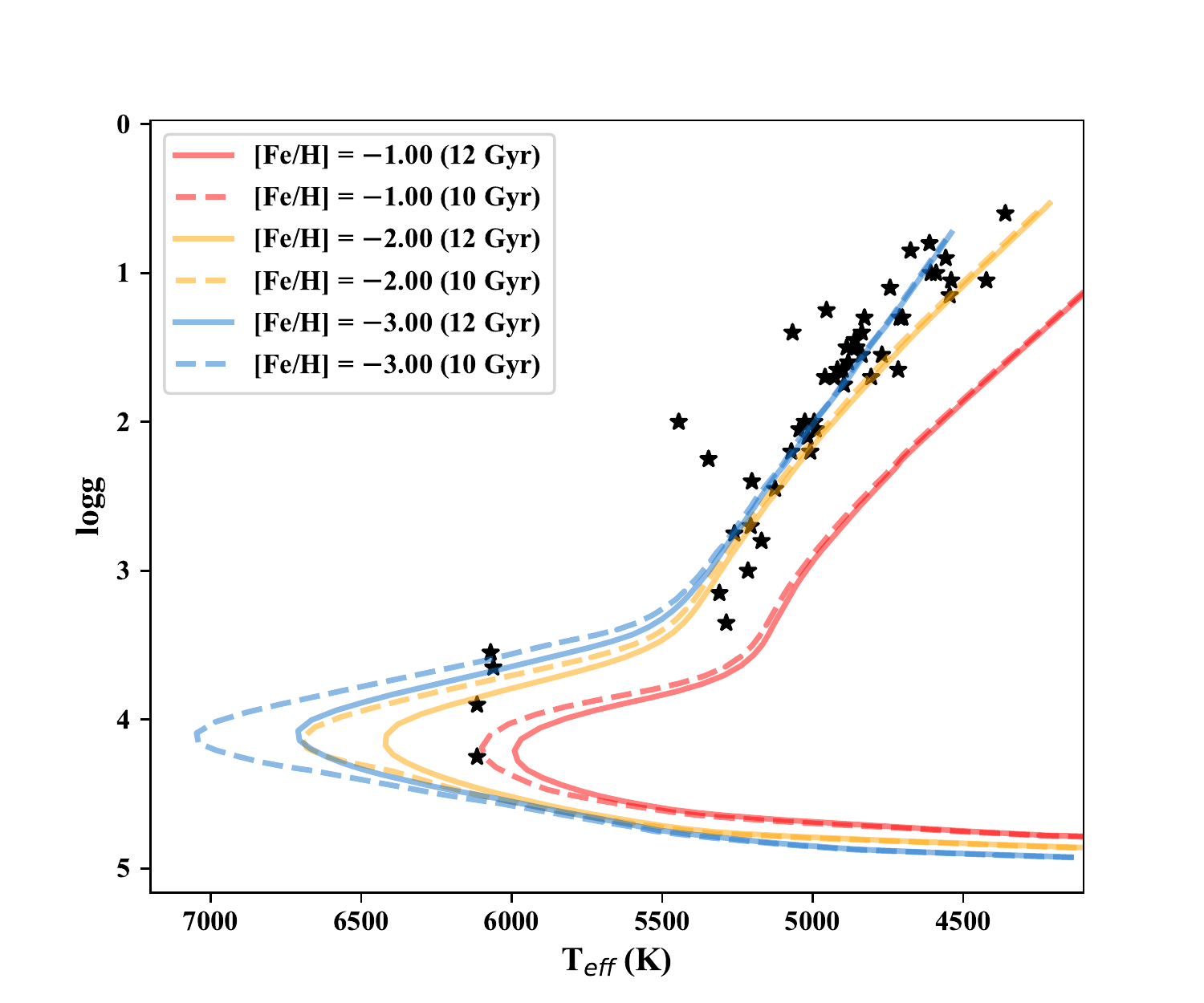}
\caption{\label{fig:iso}Comparison of the program sample to a set of isochrones. The three solid or dashed lines represent isochrones with a common age of either 10 or 12 Gyr, respectively, for metallicities of [Fe/H] = $-$1.0 (red),  $-$2.0 (orange), and $-$3.0 (blue). Data used to generate the isochrones were taken from \citet{Placco2019}.} 
\end{figure}

Effective-temperature estimates were determined by minimizing trends between the abundances of Fe\,I lines and their excitation potentials, and applying the temperature correction from the spectroscopic to the photometric scale suggested by \citet{Frebel2013}. The microturbulent velocity was determined by minimizing the trend between the abundances of Fe\,I lines and their reduced equivalent widths (REW; $\log\, ({\rm Eq. Width})/\lambda_o$). The surface gravity was determined from the balance of the two ionization stages of iron, Fe\,I and Fe\,II. The location of stars in our sample along isochrones generated for [Fe/H] = $-$1.0, $-$2.0, and $-$3.0 for both 10 and 12 Gyr, based on their derived atmospheric parameters, is shown in Figure~\ref{fig:iso}. There is little difference between the 10 and the 12 Gyr isochrones for post-main-sequence stars, but it is instructive to show that the sample matches what we expect to see for a population of old, metal-poor stars.

\subsection{Comparison with Prior Estimates}

\begin{figure}[ht!]
\includegraphics[trim={0cm 0 1cm 0}, scale=0.55]{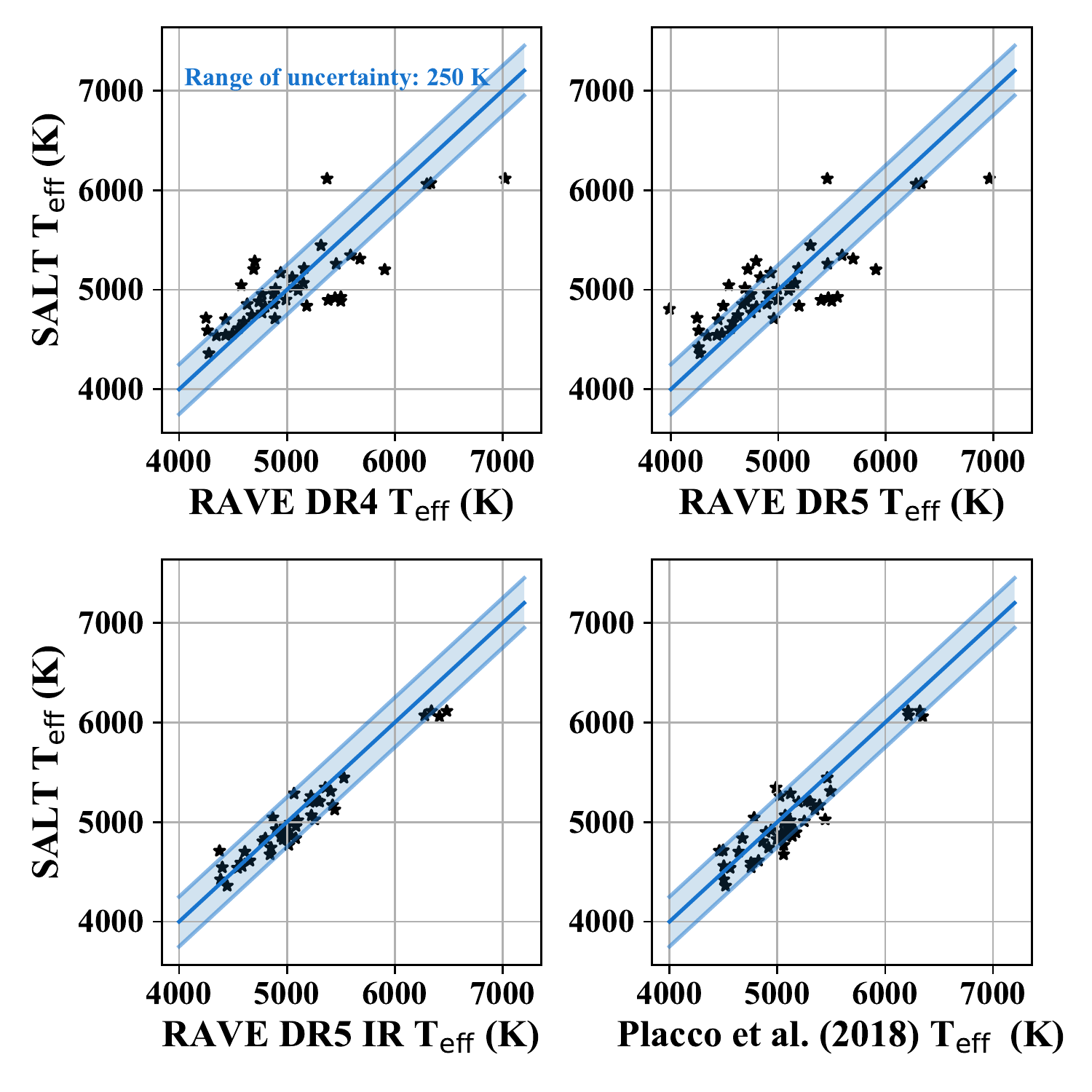}
\caption{\label{fig:teff}Temperatures from RAVE DR4 (top left), RAVE DR5 (top right), RAVE DR5 IR (bottom left), and \citet{Placco2018} (bottom right), compared with the atmospheric parameters derived in this work. The blue-shaded region marks a $\pm$250\,K margin of error. } 
\end{figure}

\textbf{Temperature:} Figure \ref{fig:teff} shows a comparison of effective-temperature estimates for our stars.  For stars with RAVE DR4 results \citep{Kordopatis2013}, there is only moderate agreement between the reported temperatures and the temperatures derived in this work (Figure~\ref{fig:teff}). RAVE DR5 \citep{Kunder2017} only issued corrections for stars above 6000\,K, which does not affect the majority of our generally cooler stars. However, new RAVE DR5 temperatures based on IR photometry exhibit far better agreement with our results, as do the temperatures derived from the n-SSPP.

\begin{figure*}[ht!]
\includegraphics[trim={4cm 0 1cm 0}, scale=0.8]{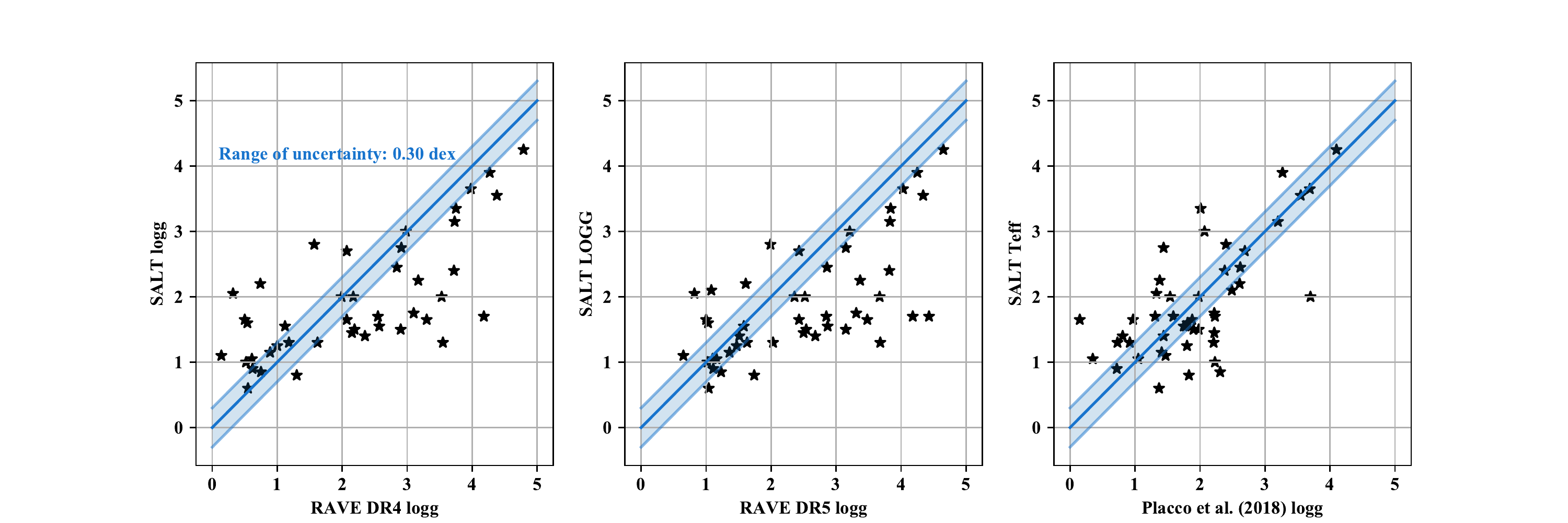}
\caption{\label{fig:logg}Log $g$ values derived from RAVE DR4 (left), RAVE DR5 (middle), and \citet{Placco2018} (right), compared with the atmospheric parameters derived in this work. The blue-shaded region marks a 0.30 dex margin of error.} 
\end{figure*}

\textbf{Surface Gravity:} Surface gravity (see Figure~\ref{fig:logg}) is often difficult to accurately estimate, even with high-resolution spectra. For both RAVE DR4 and DR5, the scatter with respect to our results is quite large, and log $g$ is frequently over-estimated by nearly a dex. Better agreement is achieved with the n-SSPP values, but the spread is still rather large.

\begin{figure*}[ht!]
\includegraphics[trim={4cm 0 1cm 0}, scale=0.8]{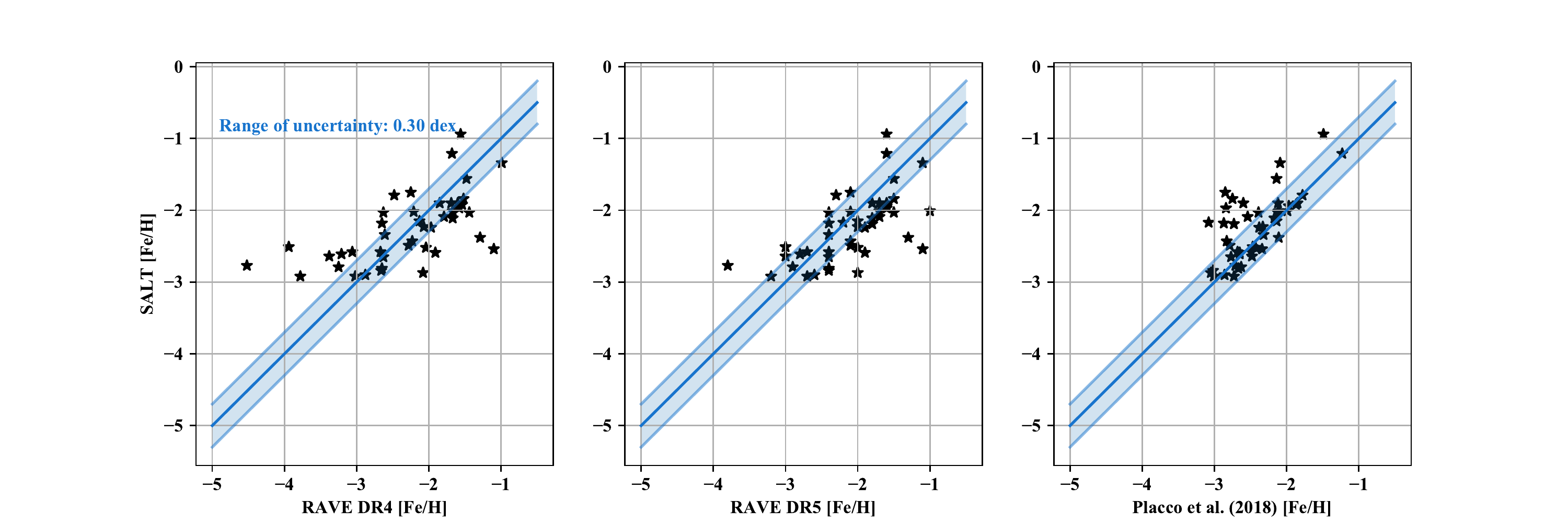}
\caption{\label{fig:feh}[Fe/H] values reported by RAVE DR4 (left), DR5 (middle), and \citet{Placco2018} (right), compared with the atmospheric parameters derived in this work. The blue-shaded region marks a 0.30 dex margin of error. Outliers in the \citet{Placco2018} comparison are generally stars which have either a very high $A$(C) value ($A$(C) $>$ 7.35), a lower temperature ($T_{\rm eff}$ $<$ 4750\,K), or both.  See text for details.} 
\end{figure*}

\textbf{Metallicity:} Figure \ref{fig:feh} shows a comparison of derived metallicities.  The metallicities from RAVE DR4 exhibit a rather large scatter compared with our SALT determinations, and include a number of highly deviant stars at the lowest [Fe/H]; the scatter and deviant behavior is reduced for RAVE DR5. The scatter compared to  the n-SSPP is generally good, but there remain a number of stars with significant deviations. Many of these outliers have either high carbon abundances ($A$(C) $>$ 7.35), lower temperature ($T_{\rm eff}$ $<$ 4750\,K), or both. This is a known difficulty for medium-resolution spectra analyzed with the n-SSPP, due in part to the nascent saturation of the CH $G$-band, and depression of the continuum by molecular carbon veiling in the region of the Ca\,II K line, which is a primary metallicity estimator for that routine. An improved methodology for dealing with both of these challenges has been recently developed, and is reported on by \citet{Yoon2020}.  

\section{Chemical Abundances}


The elements C, Mg, Ca, Sc, Ti, V, Cr, Mn, Fe, Co, Ni, Zn, Sr, Ba, and Eu have been measured (or had upper limits determined) for all 50 stars in our data set. Metallicity ranges from $\mbox{[Fe/H]} = -2.92$ to $-$0.94, with typical values around $\mbox{[Fe/H]} \sim -2.40$. Abundances of all elements, except for C, Sr, Ba, and Eu, were measured using the equivalent-width approach. Gaussian profiles were fit to all absorption lines in our line list to obtain the line measurements. 
We applied a REW cutoff of $-$4.5, as this is the point above which equivalent widths no longer linearly correlate with increasing abundance. Uncertainties for these element abundances are reported as the standard deviation resulting from multiple measurements. In the case of single-line measurements, we adopt a nominal value of 0.10\,dex. In the case of any synthesized spectral features, the uncertainty is chosen to represent the difference between the best-fit abundances and a synthetic spectrum with an abundance that encapsulates all of the data points across the region of the line. Example spectra can be found in Figure~\ref{fig:hr}. 

The derived abundances of elements useful for classifying our program stars, as discussed above, are listed in Table~\ref{tab:csreuba}. Comparison of these abundances (and the adopted atmospheric parameters) for nine of our stars with previous high-resolution spectroscopic analyses are listed in Table~\ref{tab:prev_params}. We find that our carbon and barium values values are largely consistent between works when uncertainties and differing temperatures are taken into account, and that Sr and Eu are mostly consistent, with a few outliers. For Sr, inconsistencies are likely to come from the large depth of the line combined with its many blends; for Eu, inconsistencies beyond measurement uncertainty arise when the [Eu/Fe] ratio is high and the line begins to saturate.

Additional elemental abundances for our program stars are listed in Table~\ref{tab:mg} ($\alpha$-elements), and Table~\ref{tab:v} (iron-peak elements), and are described briefly below.

\subsection{Carbon and the $\alpha$-Elements}

Carbon abundances were measured using spectral synthesis of the CH $G$-band at $\lambda$4313\,$\rm \AA$. The great majority of the stars in our sample are red giants, so we assume an equilibrium isotopic ratio value of $^{12}$C/$^{13}$C$\,\sim5.0$. These giants have experienced a varying degree of internal carbon depletion due to CN processing. To correct for this depletion, the procedure from \citet{Placco2014} was applied. Corrected C abundances are reported in Table~\ref{tab:csreuba} as [C/Fe]$_{\rm corr}$. 

Mg I abundances were measured using lines redward of $\lambda$4000\,{\AA}, including the Mg triplet, when a REW of less than $-$4.5 was reported.  Ca~I was measured in a similar fashion. No Ca~II lines were available within our wavelength range. Sc I lines, at these metallicities and S/N, and within the available wavelength range, are too weak to measure; we instead report abundances based on as many as eight Sc~II lines. We measure both Ti~I and Ti~II species, and report their abundances individually. The results for these elements are listed in Table~\ref{tab:mg}. 

\subsection{Iron-Peak Elements}

For our data, a reliable V~I abundance is reported based on only a single line at $\lambda$4786.50\,{\AA}. Both Cr~I and Cr!II abundances are reported when the S/N is sufficiently high  to measure the three weak Cr~II lines at $\lambda$4558.59\,{\AA}, $\lambda$4588.14\,{\AA}, and $\lambda$4591.99\,{\AA}. Between 60 and 80 lines were measured for Fe~I, and 5 to 15 for Fe~II, depending on the temperature, metallicity, and S/N of the stellar spectrum.

Mn~I and Ni~I abundances are measured from 7 and 18 lines, respectively. Because the majority of Co~I lines lie blueward of $\lambda$4000\,{\AA}, we derive its abundance from only the one line at $\lambda$4121.32\,{\AA}. Zn~I is measured from the two lines at $\lambda$4722.15\,{\AA} and $\lambda$4810.53\,{\AA}.  The results for these elements are reported in Table~\ref{tab:v}.  

\subsection{Neutron-Capture Elements}

Estimates of the elemental abundances for Sr, Ba, and Eu have been handled with extra care, as they are the elements by which our program metal-poor stars are classified.  Details for each element are provided below.

\textit{Strontium:} The Sr~II abundance was measured from the $\lambda$4215\,{\AA} line, as it is quite strong and does not easily over-saturate, unlike the line at $\lambda$4077\,{\AA}. The $\lambda$4161\,{\AA} line is weak and closer to the blue cutoff, so it was inaccessible in most cases.

\textit{Barium:} Ba~II was measured from the $\lambda$4934\,{\AA} line, as this feature is strong, accessible, and isolated from blends. It also does not over-saturate, as the 4554\,$\rm \AA$ is known to in cases of strong $s$-process enhancement. 

\textit{Europium}: Eu~II was measured from the $\lambda$4129\,{\AA} line, which is strong and accessible. It is blended with a Dy~II line; however, these are both $r$-process elements and reliably scale with one another over a wide range of [Eu/Fe] values. Other Eu lines exist, however, they are largely too weak to be measured at the typical S/N of our spectra.

\subsection{Comparisons with Prior Abundance Measurements}

Nine of our stars have previous estimates of stellar atmospheric parameters and chemical abundances based on from high-resolution spectroscopy, the majority of which originate from the $R$-Process Alliance data releases (\citet{Hansen2018,Sakari2018b}). These are provided in Table~\ref{tab:prev_params}. Comparing to the atmospheric parameters and abundances in the present work, we find very good agreement for several stars, while others exhibit moderate disagreements. In those cases, significant differences in v$_{\rm turb}$ can account for the differences, while for one case, our temperature estimate differs by $\sim$400\,K, with respect to that determined on a purely spectroscopic scale by \citet{Roederer2014i}; this results in large abundance differences. In addition, measurements based on relatively low- S/N spectra result in larger uncertainties for both our work and for several stars in the comparison sample. 

\section{Metal-Poor Star Sub-classifications and Frequencies}

\subsection{Classifications}

We have classified all of our program stars with astrophysically interesting chemical-abundance signatures, in order to assess their frequencies among Galactic halo stars.  Over time, as ever larger samples of metal-poor stars with these signatures become available, this should eventually enable detailed assessments of the nucleosynthetic origins of these patterns.

Sub-classifications for our program stars have been performed, based on $A$(C), [C/Fe], [Ba/Fe], and [Eu/Fe], listed along with [Sr/Fe] in Table~\ref{tab:csreuba}. This is adequate for all but two of our stars, those that are likely CEMP-$i$ stars. Such stars are difficult to classify, because confident  membership within the CEMP-$i$ class depends on a Pb abundance, a measurement we cannot obtain from our data. With higher-S/N spectra, the detailed abundance patterns could be established to further verify and fill in the details of the relevant nucleosynthesis signatures.

The distribution of CEMP stars in the $A$(C)-[Fe/H] fall into at least three groups, as defined in the Yoon-Beers diagram presented by \citet{Yoon2016}. Group I is primarily populated by relatively higher-metallicity CEMP-$s$ stars that received their high C abundances through mass transfer from a binary AGB companion.  Group II includes stars with a relatively low [Fe/H] and very low $A$(C) values, which can be classified as CEMP-no stars on the basis of either their $A$(C) or, when available, low [Ba/Fe] ratios. Group III also contains CEMP-no stars, with low [Ba/Fe], but at extremely- to ultra-low metallicities, and with higher $A$(C) than Group II CEMP-no stars.  As argued by \citet{Yoon2016}, different progenitor scenarios are thought to be responsible for the production of the stars in Groups II and III.  The availability of different cooling channels, as discussed in \citet{Chiaki2017} , may also play an important role.

\begin{figure}[ht!]
\includegraphics[trim={1cm 0 1cm 0}, scale=0.53]{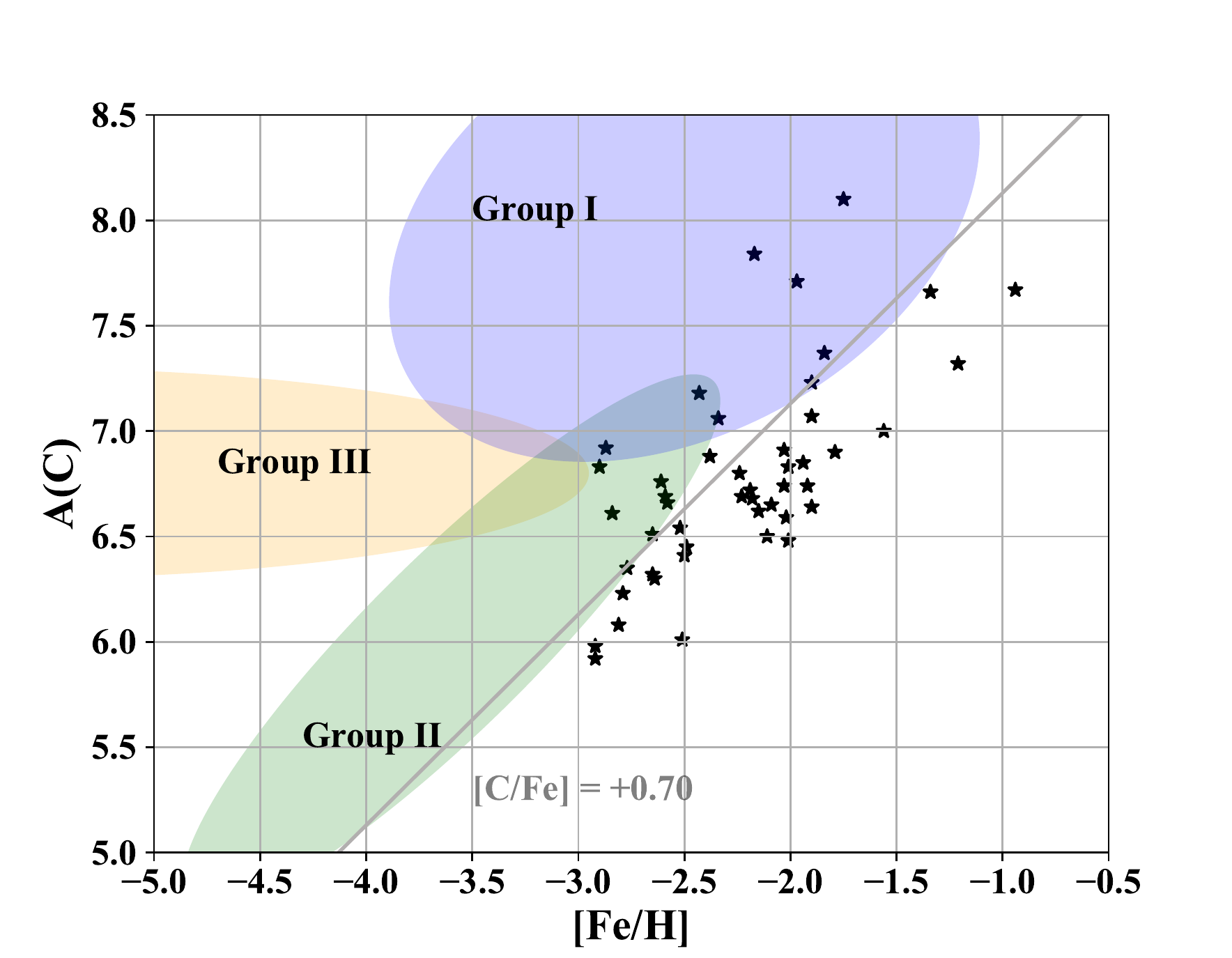}
\caption{\label{fig:yb}Location of our program stars in the $A$(C)-[Fe/H] space with overlaid groupings, following \citet{Yoon2016}. Five of the 15 CEMP stars clearly lie in Group I, seven in Group II, three in the Group I/II overlap region, and none in Group III. The rest of our sample lie below the adopted cutoff for CEMP stars,  [C/Fe] = +0.70, shown as a gray line.} 
\end{figure}

From inspection of the locations of our stars in the Yoon-Beers diagram, shown in Figure~\ref{fig:yb}, five of our CEMP stars fall into Group I, seven into Group II, and three into the region shared by Groups I and II. No stars in our sample have metallicities sufficiently low to be uniquely associated with Group III. 

\subsection{Frequencies of CEMP and $r$-Process-Enhanced Stars}

We find that 52\% of our program stars are enhanced at the [C/Fe] $>$ +0.50 level (using the corrected carbon abundances), 30\% of our sample is carbon-enhanced at the [C/Fe] $>$ +0.70 level, and 14\% of our sample  at the [C/Fe] $>$ +1.00 level. Of the 15 CEMP stars, six are CEMP-no, four are CEMP-$s$, two are likely CEMP-$i$, and three are CEMP-$r$ stars. Of the CEMP-$r$ stars, one is an $r$-I, and two are $r$-II, adding to the small number of such stars that were previously known. RAVE~J180242.3$-$440443 has [C/Fe] = +0.78 $\pm$ 0.25 with [Eu/Fe] = +1.20 $\pm$ 0.30. [Eu/Fe] has also been measured for this star by \citet{Hansen2018}, who reports [Eu/Fe] = +1.05 $\pm$ 0.30. RAVE~J171633.4$-$700902 has [C/Fe] = +0.73 $\pm$ 0.25, with [Eu/Fe] = +1.09 $\pm$ 0.30, bu no other previous measurements. A total of 10 CEMP-$r$II stars are now known, including the canonical $r$-II star CS~22892-052 (a compilation is provided in Guden et al., in prep.); it appears clear that $r$-process enrichment must at least occasionally take place in carbon-enhanced gas, presumably provided by either massive first stars in some of the earliest 
star-forming regions, or by progenitors that have been suggested to be capable of producing both C during their lifetimes and $r$-process elements when they explode, such as collapsars (see, e.g., \citealt{Siegel2019}). 

Our sample was selected to be predominantly carbon rich. Accordingly, our CEMP fraction cannot be meaningfully compared to previous literature CEMP fractions. We believe that the reason a CEMP fraction greater than 30\%  was not actually achieved, as initially intended, is likely due to a temperature offset. This can be seen in the bottom right panel of Fig~\ref{fig:teff}. On average, we arrived at temperatures 100-200\,K lower than that obtained by \citet{Placco2018} for their medium-resolution spectra. This has naturally led to lower carbon abundances for the majority of the stars in our sample. 

\textit{$r$-Process-Enhanced stars:} Half (52\%) of the program stars  of our sample are enriched in $r$-process elements, with 21 stars (42\%) being classified as $r$-I and five (10\%) classified as $r$-II. Of the $r$-II stars, three were either known previously or discovered during the course of the observing campaign by the RPA: CS~31082-001 \citep{Cayrel2001, Hill2002}, RAVE J180242.3$-$440443 \citep{Roederer2014i}), and RAVE~J153830.9$-$180424 \citep{Sakari2018a}. Two such stars stars, RAVE~J040618.2$-$030525 and RAVE~J171633.4$-$700902.8, are newly discovered, with [Eu/Fe] = +1.17 $\pm$ 0.20 and +1.09 $\pm$ 0.30, respectively. These frequencies are consistent with the RPA pilot surveys \citep{Hansen2018,Sakari2018b,Ezzeddine2020},  whose combined 374 stars contain 43\% $r$-I and 7\% $r$-II stars.

\section{Kinematics}

Kinematics for our program stars were computed using the galpy Galactic dynamics library \citep{Bovy2015}. We adopt R$_{\odot}$ = 8 kpc as the distance to the Galactic center, v$_{\rm LSR}$ = 220 km s$^{-1}$ as the local standard of rest (LSR) velocity \citep{Kerr1986}, and (U, V, W)$_{\odot}$ = ($-$9, 12, 7) km s$^{-1}$ as the motion of the Sun with respect to the LSR \citep{Mihalas1981}. Orbital parameters were derived with the potential code employed by \citet{Chiba2000}, which utilizes the  St\"ackel potential described in \citet{Sommer-Larsen1990}. Uncertainties in the orbital parameters were derived through a Markov chain Monte Carlo (MCMC) sampling method.

\begin{figure}[ht!]
\includegraphics[trim={0.5cm 0 1cm 0}, scale=0.57]{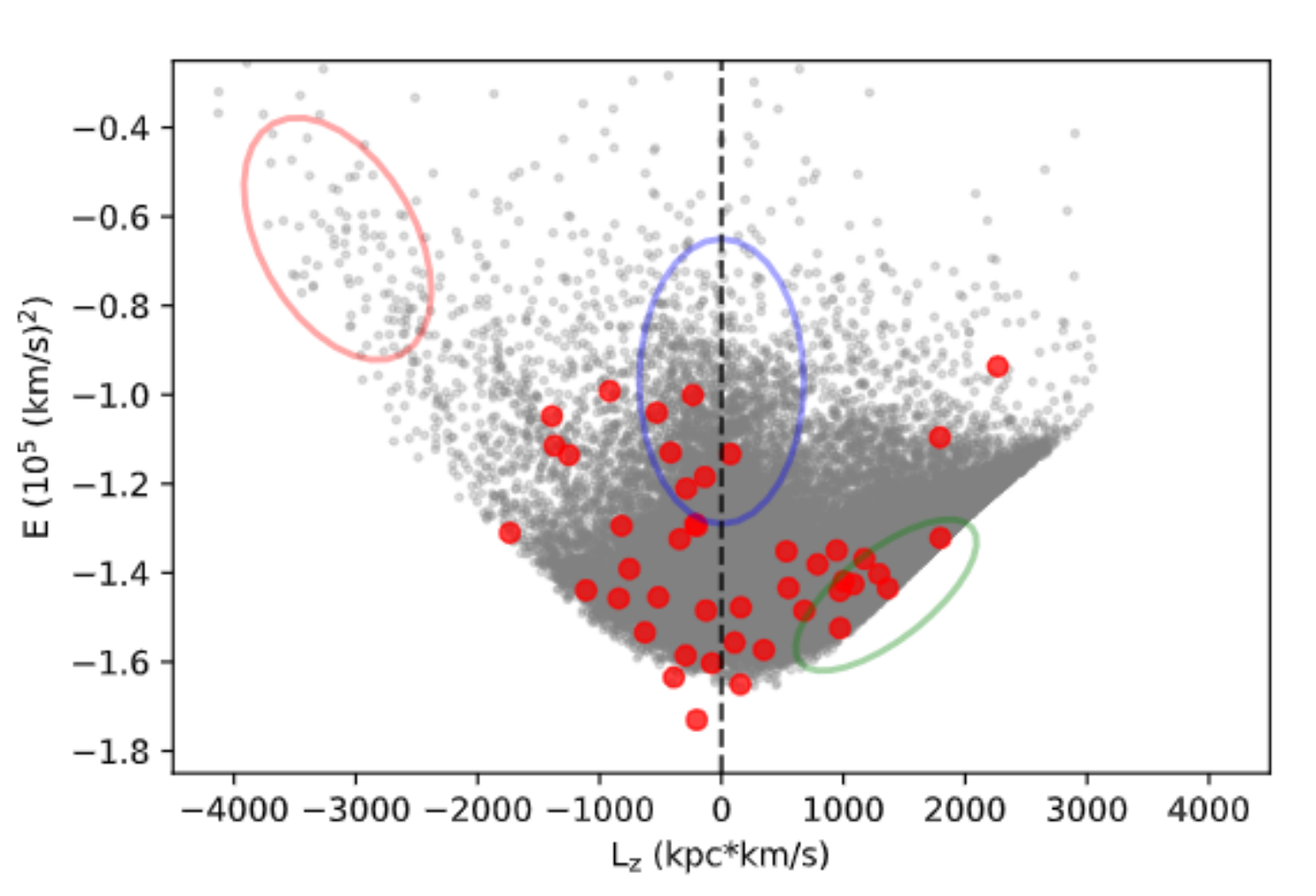}
\caption{\label{fig:kin} Stars in our sample (red filled circles) plotted in Energy-L$_{\rm Z}$ space on top of a reference sample of SDSS stars (gray dots) from Dietz et al. (in prep). The dashed line marks the L$_{\rm z}$ = 0 line, separating prograde and retrograde populations. The green ellipse indicates the strongly-bound, prograde disk population, blue indicates the ``plume" that marks the remnant of the large inner-halo progenitor referred to as $Gaia$-Enceladus \citep{Helmi2018} or the $Gaia$-Sausage \citep{Belokurov2018,Myeong2018}, and pink indicates high-energy, retrograde stars in the outer halo.}
\end{figure}

To better represent where our sample stars lie in the Lindblad diagram with respect to various Galactic populations, Figure~\ref{fig:kin} over-plots our stars with a reference sample from SDSS (Dietz et al. in prep.). The red filled dots indicate our program stars. The dashed line marks the L$_{\rm z}$ = 0 line, separating prograde and retrograde populations. The green, blue, and red ellipses indicate approximate locations of important reference populations. Green indicates the strongly bound, prograde disk system. Blue indicates the ``plume” that marks the remnant of the large inner-halo progenitor referred to as Gaia Enceladus \citep{Helmi2018} or the Gaia Sausage \citep{Belokurov2018,Myeong2018}. Red indicates high-energy, retrograde stars in the outer halo. The stars in our sample, being relatively bright and (presently) close-by, are largely more strongly bound, and comprise members of the disk system and inner-halo population. No high-energy outer-halo stars are found in our sample,
although the more energetic stars have higher uncertainties on their orbital-value determinations, so their membership is somewhat uncertain.

\begin{figure*}[ht!]
\centering
\includegraphics[trim={1cm 0 1cm 0}, scale=0.67]{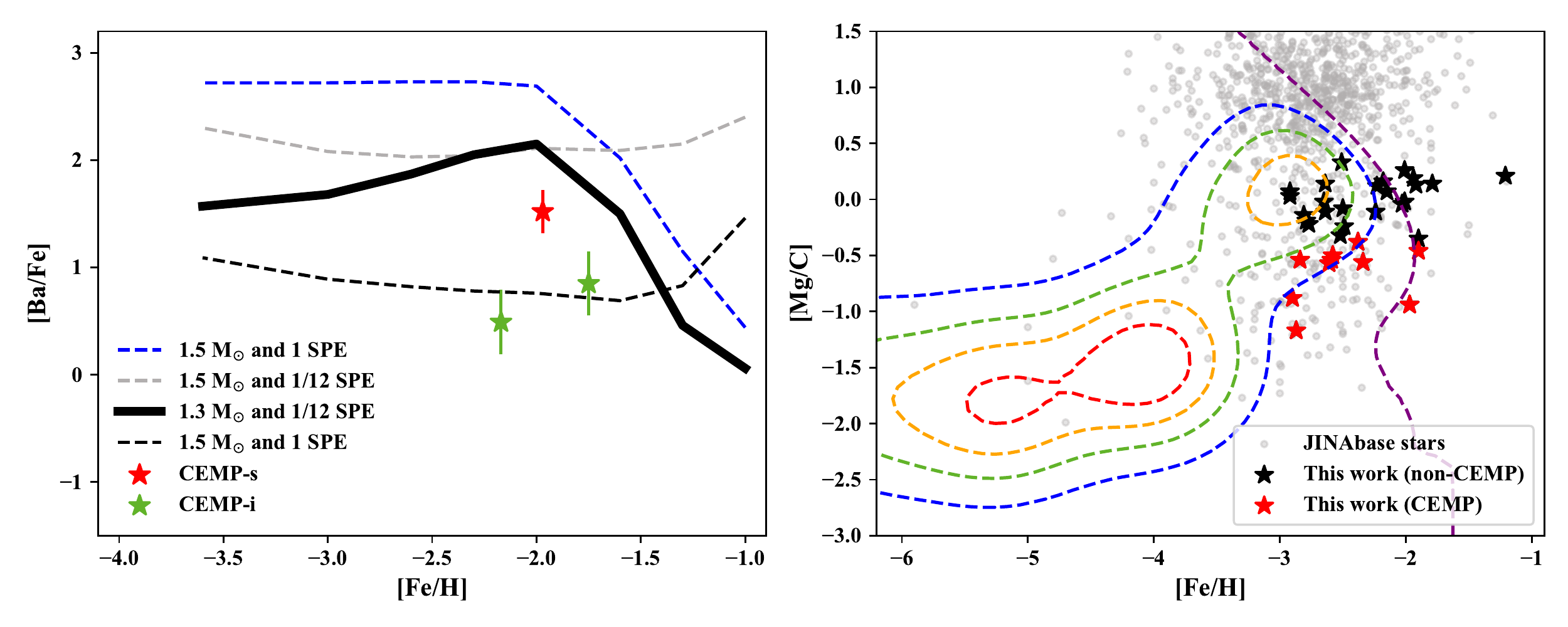}
\caption{\label{fig:hartwig}(Left Panel) The three Ba-rich stars in our sample (the CEMP-$s$ star RAVE~J111912.3-160947, and the CEMP-$i$ stars RAVE~J044208.2$-$342114 and  RAVE~J201446.1$-$563530), are plotted in the [Ba/Fe]-[Fe/H] space, along with theoretical yield predictions for two AGB stars with different initial masses and $^{13}$C-pocket efficiencies, where the C-pocket refers to the thin layer of radiative carbon burning that occurs in the top layer of the He shell after the third dredge-up episode of an AGB star. The dashed gray and blue lines represent models with 1.5 M$_{\odot}$ and 1/12 and 1 standard pocket efficiency (SPE), respectively. The solid black and dashed black lines represent models with 1.3 M$_{\odot}$ and 1/12 SPE and 1 SPE, respectively. Data were taken from Tables B5 and B6 of \citet{Bisterzo2010}. We find that our CEMP-$i$ stars fall along the 1.3 M$_{\odot}$ SPE track, while the CEMP-$s$ star falls between the 1.3 M$_{\odot}$ 1/12 SPE and 1 SPE track. (Right Panel) Comparison of our newly identified stars with available C and Mg abundances to a diagnostic for mono-enrichment or multi-enrichment proposed by \citet{Hartwig2018}. The contours illustrate the likelihood for a star to be mono-enriched, based on its chemical composition. The majority of stars appear consistent with mono-enrichment, with only a small fraction, located to the right of the purple contour, likely being enriched by more than one supernova. } 
\end{figure*}

\section{Comparison with $s$- and $i$-Process Models and Origin Scenarios}

Finally, we compare our results with chemical-abundance models in order to investigate the $^{13}$C-pocket efficiencies (where the C-pocket refers to the thin layer of radiative carbon burning that occurs in the top layer of the He-shell after the third dredge-up episode of an AGB star) ultimately responsible for our three Ba-rich stars. We also consider the divergence of chemical displacement scheme to investigate origin scenarios of our program stars, by testing whether they formed from gas enriched by multiple vs. single supernovae.

\subsection {The Ba-Rich Stars}

Three of our program CEMP stars are highly enriched in Ba, likely due to mass transfer from an AGB companion -- one CEMP-$s$ star (RAVE~J111912.3-160947) and two CEMP-$i$ stars (RAVE~044208.2-342114, RAVE~201446.1-563530).  All three are compared with theoretical predictions for a variety of AGB models, varying in mass and $^{13}$C-pocket efficiencies, taken from \citet{Bisterzo2010}. We find that a model of 1.3\,M$_{\odot}$ with standard pocket efficiency fits our CEMP-$i$ stars well, but that a lower efficiency may be required to fit the abundances of the CEMP-$s$ star. The results are shown in the left panel of Figure~\ref{fig:hartwig}. Similar tests for a larger sample of CEMP stars from our full $\sim$200 star sample will be considered once their analysis is completed.

\subsection{Origin Scenarios}

We consider the origin scenarios for our program stars, contrasting the cases of enrichment by a single (mono-enriched) supernova with that from several supernovae (milti-enriched), using the divergence of the chemical displacement approach described by \citet{Hartwig2018}, and shown in the right panel of Figure~\ref{fig:hartwig}. Broadly, stars that are mono-enriched are expected to occupy a larger region in the [Mg/C] vs. [Fe/H] abundance space. For stars that form from gas enriched by multiple supernovae, this region shrinks, since the abundances of such stars are effectively weighted averages of mono-enriched stars, resulting in a more centrally concentrated space. This change in the occupation of abundance space can be expressed by a vector field of chemical displacement, and its divergence quantifies regions of mono- and multi-enrichment. The contours illustrate a positive divergence (from 0 to 160 in steps of 40; compare with Figure~14 in \citealt{Hartwig2018}), and highlight regions of the abundance space where we expect stars to be mono- vs. multi-enriched.

Only two out of nine newly discovered CEMP stars for which Mg abundances are available exhibit a mild tendency for multi-enrichment (the region to the right of the purple contour in  Figure~\ref{fig:hartwig}, right panel). Their location on the this diagram is driven by their relatively high [Fe/H] ([Fe/H] $\sim-$1.9). One is a strongly bound CEMP-no star on a retrograde orbit (J111711.0$-$310951), and one is a CEMP-$s$ star (J111912.3$-$160947) that lies very close to the inner-halo population in the Energy-L$_{z}$ space shown in Figure~\ref{fig:kin}. The other stars, which are more metal poor (down to [Fe/H] $\sim-$2.9), follow along the blue contour indicating mono-enrichment; they do not preferentially occupy any particular location in Figure~\ref{fig:kin}. This implies that the majority of the CEMP stars in our sample are likely to have formed from gas enriched by only one progenitor supernova, rather than by many. These results are consistent with expectations based on yields from faint supernovae by \citet{Nomoto2013} and \citet{Ishigaki2014}. 


\section{Summary \& Outlook}

We present a limited set of chemical-abundance measurements and kinematics for a new sample of 50 metal-poor (MP; [Fe/H] $< -1.0$) and very metal-poor (VMP; [Fe/H] $< -2.0$) stars in the metallicity range $-$2.92 $<$ [Fe/H] $<$ $-$0.94, based on high-resolution ($R \sim 40,000$ spectroscopic data collected with the HRS on the Southern African Large Telescope. Sub-classification of our program stars indicate that 15 (30\%) of these are CEMP stars; nine of these are CEMP-no stars (eight of which are newly discovered), one is a CEMP-$s$ star, and two are likely CEMP-i stars.  Twenty-six  of our program stars (52\%) are $r$-process-enhanced stars (21 $r$-I, and newly discovered, and 5 $r$-II stars, two of which are newly discovered); both of these latter stars are newly discovered members of the sparsely populated CEMP-$r$ sub-class.

We also find that there are eight stars in our sample that, on the basis of their abundances and kinematics, are possible members of the metal-weak thick disk. Further chemical-abundance analysis at higher resolution over an expanded wavelength range would prove useful in the study of the differing enrichment mechanisms which are responsible for the differences in the chemical-evolution histories of the halo and the thick disk.

We compare our stellar parameters with those obtained by RAVE DR4, RAVE DR5, and \citet{Placco2018}, and find moderate to good agreement for most stars.  We argue that a number of the stars with [Fe/H] and [C/Fe] estimates from SALT disagree with those reported by \citet{Placco2018}, most likely due to their cooler temperatures and relatively strong molecular carbon bands.  Application of a new procedure to the medium-resolution spectra for these stars, described by \citet{Yoon2020}, results in improved agreement with our high-resolution SALT determinations. Comparison with previous stellar atmospheric parameters and a limited set of chemical abundances with previous high-resolution analyses for nine stars indicate reasonably good agreement. 

We also compare our results with chemical-enrichment models to investigate the $^{13}$C-pocket efficiencies of AGB stars that are likely responsible for production of the heavy elements for our three Ba-rich stars, and show that most of our program stars are consistent with mono-, rather than multi-enrichment, scenarios.   

Additional higher S/N high-resolution spectra for the four stars in this sample belonging to poorly populated sub-classes (two CEMP-$r$II stars and two CEMP-$i$ stars), from which a larger number of neutron-capture elements cam be obtained, would be especially valuable.  Analysis of the remaining $\sim$150 stars from our SALT program, now underway (Zepeda et al., in prep.) will improve our estimates of the frequencies of the various sub-classes found in the present sample, and identify additional astrophysically interesting stars. Long-term radial-velocity monitoring for many of these relative bright MP and VMP stars will also prove illuminating.

\acknowledgments{
The authors thank an anonymous referee for positive comments and suggestions that improved the final paper.  K.C.R., J.Z., T.C.B., V.M.P., A.F., and S.D. acknowledge partial support from grant PHY 14-30152; Physics Frontier Center/JINA Center for the Evolution of the Elements (JINA-CEE),
awarded by the US National Science Foundation.  A.F. acknowledges partial support from NSF grant AST-1716251.}

\facilities {\textit{Southern African Large Telescope (SALT)} High Resolution Spectrograph (HRS), \textit{Mayall 4m Telescope}, \textit{ESO New Technology Telescope}. }
\\

\clearpage
\newpage

\bibliographystyle{aasjournal}
\bibliography{library2.bib}





\begin{deluxetable}{lcccr}
\tablecaption{\label{tab:obs}{}Details of the Observations}
\tabletypesize{\scriptsize}
\tablehead{
\colhead{Name}  & 
\colhead{Obs. Date} &
\colhead{Exp. Time (hrs)} &
\colhead{S/N @ 4310\,\AA} &
\colhead{V}}

\startdata
RAVE J012931.1$-$160046$^{a}$ & 2017 Jun 11 & 2.65 & 23 & 11.64\\
RAVE J030639.1$-$692040 & 2018 Jul 21 & 1.75 & 15 & 12.36\\
RAVE J035550.9$-$721151 &  2017 Jun 13 & 1.95 & 20 & 13.35\\
RAVE J040618.2$-$030525 & 2017 Oct 28 & 1.80 & 19 & 12.17\\
RAVE J044208.2$-$342114$^{b}$ &  2017 Oct 26 & 2.25 & 18 & 11.65\\
RAVE J045322.6$-$221040$^{c}$ & 2017 Oct 26 & 1.60 & 25 & 10.66\\
RAVE J053817.0$-$751621 &  2017 Nov 05 & 2.20 & 21 & 11.21\\
RAVE J061950.0$-$531212 &  2017 Oct 26 & 1.90 & 18 & 11.37\\
\enddata
\tablecomments{
 \mbox{A handful of these stars} have been previously observed in other programs since the beginning of the observing campaign or earlier and are noted as such: (a) \citet{Hansen2015}, (b) \citet{Sakari2018a}, (c) \citet{Sakari2018b} (d) \citet{Hansen2018}, (e) \citet{Roederer2014i}. Their stellar atmospheric parameters and [C/Fe], [Sr/Fe], [Ba/Fe], and [Eu/Fe] values are compared in Table~\ref{tab:prev_params}.}
\end{deluxetable}

\begin{deluxetable*}{lllccccccccc}
\tablecaption{\label{tab:prev_params}Stellar Atmospheric Parameters and Abundances for Previously Observed Stars}
\tabletypesize{\scriptsize}
\tablehead{
\colhead{Name}  & 
\colhead{Other Name} & 
\colhead{Lit.} & 
\multicolumn{4}{c}{Literature Values} &
\colhead{} &
\multicolumn{4}{c}{This Work}\\
\cline{4-7} \cline{9-12}
\colhead{} &
\colhead{} &
\colhead{} &
\colhead{$T_{\rm eff}$ \, (K)} & 
\colhead{log $g$}\,(cgs) &
\colhead{[Fe/H]} & 
\colhead{v$_{\rm turb}$}\,(km s$^{-1}$) & 
\colhead{} &
\colhead{$T_{\rm eff}$}\,(K) & 
\colhead{log $g$}\,(cgs) & 
\colhead{[Fe/H]} & 
\colhead{v$_{\rm turb}$}\,(km s$^{-1}$)
}
\startdata
RAVE J012931.1$-$160046 & BPS CS 31082-001 & 1 & 4800 & 1.50 & $-$3.01 & 1.95 & & 4959 & 1.70 & $-$2.77 & 2.65 \\
RAVE J044208.2$-$342114 & HE 0440-3426 & 2 & 4800 & 1.60 & $-$2.20 & 1.90 & & 4837 & 1.40 & $-$2.17 & 2.25 \\
RAVE J045322.6$-$221040 & TYC 5911-452-1 & 3 & 6225 & 4.54 & $-$1.88 & 2.33 & & 6115 & 3.90 & $-$2.54 & 1.60 \\
RAVE J153830.9$-$180424 & TYC 6189-285-1 & 4 & 4752 & 1.63 & $-$2.09 & 1.51 & & 4995 & 2.00 & $-$2.02 & 1.85 \\ 
RAVE J160245.0$-$152102 & BD $-$14 4336 & 5 & 5240 & 3.16 & $-$1.80 & 1.65 & & 5215 & 3.00 & $-$1.90 & 1.50 \\
RAVE J180242.3$-$440443 & UCAC2 13007354 & 5 & 4701 & 1.60 & $-$1.55 & 2.17 & & 4711 & 1.30 & $-$1.84 & 2.30\\
RAVE J205609.1$-$133118 & HD 358059 & 5 & 4780 & 1.60 & $-$2.30 & 1.80 & & 4918 & 1.65 & $-$2.24 & 1.95 \\
RAVE J210642.9$-$682827 & UCAC2 2203606 & 5 & 5186 & 2.70 & $-$2.76 & 2.90 & & 5026 & 2.00 & $-$2.92 & 1.55 \\
\enddata

\tablenotetext{a}{The [C/Fe] value reported is ``as measured", not corrected for evolutionary effects, as in the rest of the paper.}
\tablecomments{Literature citations are as follows: 1) \citet{Hill2002} 2) \citet{Hansen2015} 3) \citet{Sakari2018a} 4) \citet{Sakari2018b} 5) \citet{Hansen2018} 6) \citet{Roederer2014i}. \mbox{Average} Uncertainties for [X/Fe] measurements are as follows: \citet{Hill2002}: $\sim$0.05-0.20\,dex; \citet{Hansen2015}: $\sim$0.2-0.3\,dex; \citet{Sakari2018b}: $\sim$0.05-0.2\,dex; \citet{Sakari2018a}: $\sim$0.1-0.2\,dex; \citet{Hansen2018}: $\sim$0.3-0.4\,dex; \citet{Roederer2014i}: $\sim$0.2-0.3\,dex.}
\end{deluxetable*}

\begin{deluxetable*}{lrcrc}
\tablecaption{\label{tab:params}Stellar Atmospheric Parameters for the Program Stars}
\tabletypesize{\scriptsize}
\tablehead{
\colhead{Name}  & 
\colhead{$T_{\rm eff}$\,(K)} & 
\colhead{log $g$\,(cgs)} &
\colhead{[Fe/H]} &
\colhead{v$_{\rm turb}$\,(km s$^{-1}$)}
}
\startdata
RAVE J012931.1$-$160046&4959 & 1.70 & $-$2.77 & 2.65 \\
RAVE J030639.1$-$692040&4770 & 1.55 & $-$2.81 & 1.75 \\
RAVE J035550.9$-$721151&4990 & 2.05 & $-$2.64 & 1.95 \\
RAVE J040618.2$-$030525&5260 & 2.75 & $-$1.34 & 1.80 \\
RAVE J044208.2$-$342114&4837 & 1.40 & $-$2.17 & 2.25 \\
RAVE J045322.6$-$221040&6115 & 3.90 & $-$2.54 & 1.60 \\
RAVE J053817.0$-$751621&4900 & 1.65 & $-$2.03 & 2.20 \\
RAVE J061950.0$-$531212&5170 & 2.80 & $-$2.15 & 1.90 \\
\enddata
\end{deluxetable*}

\begin{deluxetable*}{lrrccccc}
\tablecaption{\label{tab:csreuba}Classification Abundances for the Program Stars}
\tabletypesize{\scriptsize}
\tablehead{
\colhead{Name}  & 
\colhead{Class} & 
\colhead{$A$(C)$_{\rm corr}$} &
\colhead{[C/Fe]} &
\colhead{[C/Fe]$_{\rm corr}$} &
\colhead{[Sr/Fe]} &
\colhead{[Ba/Fe]} & 
\colhead{[Eu/Fe]}}
\startdata
RAVE J012931.1$-$160046 & $r$-II & 6.35 & +0.44 $\pm$ 0.25 & +0.69 $\pm$ 0.25 & +0.96 $\pm$ 0.30 & +1.20 $\pm$ 0.30 & +1.80 $\pm$ 0.30 \\
RAVE J030639.1$-$692040 & \dots & 6.08 & +0.08 $\pm$ 0.25 & +0.46 $\pm$ 0.25 & $-$0.53 $\pm$ 0.30 & $-$0.31 $\pm$ 0.20 & $<$+0.35     \\
RAVE J035550.9$-$721151 & $r$-I &  6.11   & +0.32 $\pm$ 0.25 & +0.32  $\pm$ 0.25 & +0.53 $\pm$ 0.20 & $-$0.08 $\pm$ 0.2 & +0.60 $\pm$ 0.20 \\
RAVE J040618.2$-$030525 & $r$-II & 7.66 & +0.54 $\pm$ 0.25 & +0.57 $\pm$ 0.25 & \dots            & +0.98 $\pm$ 0.30 & +1.17 $\pm$ 0.20 \\
RAVE J044208.2$-$342114 &  CEMP-$i$ & 7.84 & +1.44 $\pm$ 0.25 & +1.58 $\pm$ 0.25 &       \dots      & +0.49 $\pm$ 0.30 & +0.39 $\pm$ 0.2 \\
RAVE J045322.6$-$221040 & \dots & $<$ 6.19  & $<$ 0.00 &    $<$+0.30         & $-$0.29 $\pm$ 0.10 & $-$0.73 $\pm$ 0.10 & \dots        \\
RAVE J053817.0$-$751621 & \dots & 6.91 & +0.18 $\pm$ 0.25 & +0.51 $\pm$ 0.25 & $-$0.63 $\pm$ 0.30 & $-$0.24 $\pm$ 0.20 & +0.44 $\pm$ 0.20 \\
RAVE J061950.0$-$531212 & \dots & 6.62 & +0.33 $\pm$ 0.25 & +0.34 $\pm$ 0.20 & +1.24 $\pm$ 0.20 & $-$0.24 $\pm$ 0.20 & +0.51 $\pm$ 0.20 \\
\enddata

\end{deluxetable*}

\begin{deluxetable*}{lrccccc}
\tablecaption{\label{tab:mg} Abundance Ratios for the Elements Mg, Ca, Sc II, Ti I, and Ti II}
\tabletypesize{\scriptsize}
\tablehead{
\colhead{Name}  & 
\colhead{[Mg/Fe]} & 
\colhead{[Ca/Fe]} &
\colhead{[Sc II/Fe]} &
\colhead{[Ti I/Fe]} & 
\colhead{[Ti II/Fe]}}

\startdata
RAVE J012931.1$-$160046 & +0.47 $\pm$ 0.19 & +0.47 $\pm$ 0.15 & +0.03 $\pm$ 0.01 & +0.31 $\pm$ 0.18 & +0.33 $\pm$ 0.12 &  \\
RAVE J030639.1$-$692040 & +0.32 $\pm$ 0.10 & \dots            & $-$0.10 $\pm$ 0.10 & +0.14 $\pm$ 0.10 & +0.12 $\pm$ 0.08 &  \\
RAVE J035550.9$-$721151 & +0.46 $\pm$ 0.10 & +0.42 $\pm$ 0.14 & +0.11 $\pm$ 0.03 & +0.22 $\pm$ 0.07 & +0.37 $\pm$ 0.11 &  \\
RAVE J040618.2$-$030525 & \dots            & +0.41 $\pm$ 0.05 & +0.47 $\pm$ 0.10 & +0.23 $\pm$ 0.10 & +0.27 $\pm$ 0.15 &  \\
RAVE J044208.2$-$342114 & \dots            & +0.48 $\pm$ 0.04 & +0.24 $\pm$ 0.10 & +0.26 $\pm$ 0.11 & +0.23 $\pm$ 0.08 &  \\
RAVE J045322.6$-$221040 & +0.27 $\pm$ 0.04 & +0.44 $\pm$ 0.10 & +0.21 $\pm$ 0.07 & +0.43 $\pm$ 0.06 & +0.49 $\pm$ 0.09 &  \\
RAVE J053817.0$-$751621 & \dots            & +0.60 $\pm$ 0.22 & +0.05 $\pm$ 0.09 & +0.24 $\pm$ 0.14 & +0.31 $\pm$ 0.15 &  \\
RAVE J061950.0$-$531212 & +0.41 $\pm$ 0.03 & +0.33 $\pm$ 0.11 & +0.04 $\pm$ 0.05 & +0.25 $\pm$ 0.11 & +0.33 $\pm$ 0.12 &  \\
\enddata
\tablecomments{\mbox{Uncertainties are taken} as the standard deviation of multiple measurements; in cases where only one line was measurable, a nominal uncertainty of 0.10\,dex has been assigned.}
\end{deluxetable*}

\begin{deluxetable*}{lccccccc}
\caption{\label{tab:v} Abundance Ratios for the Elements V I, Cr I, Cr II, Mn, Co, Ni, and Zn}
\tabletypesize{\scriptsize}
\tablehead{
\colhead{Name}  & 
\colhead{[V I/Fe]} & 
\colhead{[Cr I/Fe]}  & 
\colhead{[Cr II/Fe]} & 
\colhead{[Mn/Fe]} &
\colhead{[Co/Fe]} &
\colhead{[Ni/Fe]} & 
\colhead{[Zn/Fe]}}
\startdata
RAVE J012931.1$-$160046 &       \dots      & $-$0.21 $\pm$ 0.16 &     \dots        & $-$0.44 $\pm$ 0.09 & +0.64 $\pm$ 0.10 & +0.01 $\pm$ 0.12 & +0.30 $\pm$ 0.10   \\
RAVE J030639.1$-$692040 &       \dots      & $-$0.32 $\pm$ 0.11 &      \dots       & $-$0.47 $\pm$ 0.10 & -0.09 $\pm$ 0.10  & +0.14 $\pm$ 0.16 &     \dots          \\
RAVE J035550.9$-$721151 & +0.13 $\pm$ 0.10 & $-$0.23 $\pm$ 0.08 & +0.18 $\pm$ 0.02 & $-$0.41 $\pm$ 0.10 & +0.48 $\pm$ 0.02 & +0.11 $\pm$ 0.06 & +0.26 $\pm$ 0.01   \\
RAVE J040618.2$-$030525 & +0.29 $\pm$ 0.10 & $-$0.03 $\pm$ 0.14 & -0.09 $\pm$ 0.10 & $-$0.23 $\pm$ 0.03 &      \dots       & +0.00 $\pm$ 0.08 & $-$0.01 $\pm$ 0.13   \\
RAVE J044208.2$-$342114 &   \dots & +0.00 $\pm$ 0.21 &       \dots      & $-$0.17 $\pm$ 0.11 &       \dots      & +0.00 $\pm$ 0.06 & +0.10 $\pm$ 0.10   \\
RAVE J045322.6$-$221040 &       \dots      & $-$0.17 $\pm$ 0.14 &      \dots       & $-$0.44 $\pm$ 0.10 & +0.17 $\pm$ 0.10 & +0.25 $\pm$ 0.10 &       \dots        \\
RAVE J053817.0$-$751621 & +0.06 $\pm$ 0.10 & $-$0.11 $\pm$ 0.10 &     \dots        & $-$0.35 $\pm$ 0.05 & -0.12 $\pm$ 0.10 & +0.05 $\pm$ 0.11 & +0.01 $\pm$ 0.03   \\
RAVE J061950.0$-$531212 &  \dots & $-$0.20 $\pm$ 0.09 &      \dots       & $-$0.49 $\pm$ 0.07 & +0.54 $\pm$ 0.08 & +0.11 $\pm$ 0.05 & +0.44 $\pm$ 0.10  \\
\enddata
\tablecomments{Uncertainties are taken as the standard deviation of multiple measurements; in cases where only one line was measurable, a nominal uncertainty of 0.10\,dex has been assigned.}
\end{deluxetable*}

\clearpage

\section{Appendix}

Here we provide a table of the atomic data for lines used in the equivalent-width analysis of the SALT spectra.

\startlongtable

\begin{deluxetable*}{lrrr}
\tablewidth{5in}
\tablecaption{\label{tab:lines}Line List for Abundances}
\tablehead{
\colhead{Ion}  & 
\colhead{$\lambda$} & 
\colhead{$\chi$} &
\colhead{log $gf$}}
\startdata
          &           &           &             \\
     Mg I &  4057.505 &      4.35 &    $-$0.890 \\
     Mg I &  4167.271 &      4.35 &    $-$0.710 \\
     Mg I &  4351.906 &      4.34 &    $-$0.525 \\
     Mg I &  4571.096 &      0.00 &    $-$5.688 \\
     Mg I &  4702.990 &      4.33 &    $-$0.380 \\
     Mg I &  5172.684 &      2.71 &    $-$0.450 \\
     Mg I &  5183.604 &      2.72 &    $-$0.239 \\
\enddata 
\end{deluxetable*}

\end{document}